\documentclass[aps,preprint,nofootinbib,superscriptaddress]{revtex4}
\usepackage{amssymb}
\usepackage{graphicx}
\usepackage{float}
\begin{document}

\title {The Study of Mass Distribution of products in 7.0 AMeV
$^{238}$U+$^{238}$U Collisions}

\author{Kai Zhao}
\affiliation{China Institute of Atomic Energy, P.O. Box 275(18),
Beijing 102413, P. R. China}
\author{Xizhen Wu}%
\email{lizwux9@ciae.ac.cn} \affiliation{China Institute of Atomic
Energy, P.O. Box 275(18), Beijing 102413, P. R. China}
\author{Zhuxia Li}
\email{lizwux@ciae.ac.cn} \affiliation{China Institute of Atomic
Energy, P.O. Box 275(18), Beijing 102413, P. R. China} \affiliation{
Institute of Theoretical Physics, Chinese Academic of Science,
Beijing 100080, P. R. China}

\date{\today}
\

\begin{abstract}
Within the Improved Quantum Molecular Dynamics (ImQMD) Model
incorporating the statistical decay Model, the reactions of
$^{238}$U+$^{238}$U at the energy of 7.0 AMeV have been studied.
The charge, mass and excitation energy distributions of primary
fragments are investigated within the ImQMD model and
de-excitation processes of those primary fragments are described
by the statistical decay model. The mass distribution of the final
products in $^{238}$U+$^{238}$U collisions is obtained and
compared with the recent experimental data.
\end{abstract}

\maketitle

\begin{center}
\bf{I. Introduction}
\end{center}

When beams in the actinide region with bombarding energies above
the Coulomb barrier became available about twenty years ago, the
strongly damped reactions in very heavy systems, such as in
$^{238}U$ + $^{238}U$ were
studied\cite{Hild1065,Schadel41,Freies292}. These early
experiments emphasized the investigation on the decay channels of
the di-nuclear system (for production of superheavy nuclei) or on
particle creation in the strong electromagnetic fields. Recently,
renewed interest in this subject has been motivated by the
necessity of clarifying the dynamics of very heavy nuclear
collisions at low excitation energies and by the search for new
ways of producing neutron-rich superheavy nuclei. Based on coupled
Langevin-type equations, a model for the simultaneous description
of deep inelastic scattering, quasi-fission, fusion and regular
fission was proposed in ref. \cite{Zagrebaev34}. Within this model
the reactions of $^{238}$U+$^{238}$U, $^{232}$Th+$^{250}$Cf and
$^{238}$U+$^{248}$Cm were investigated and a large transfer of
charge and mass were found in those reactions as a result of an
inverse quasi-fission process\cite{Zagrebaev06,Zagrebaev34}. Owing
to very heavy nuclear system and very complicated process, a large
number of degrees of freedom, such as the excitation and
deformation of projectile and target, the neck formation, nucleon
transfer, different types of separation of the composite system
and nucleon emission will simultaneously play a role. Thus, one
faces a difficulty for handling the problem with such complex
mechanism and large number of degrees of freedom by the
macroscopic dynamics model. In this case, a microscopic transport
theory model is worthy to be used\cite{Wang05,Tian08}. In ref.
\cite{Tian08} the formation and properties of the transiently
formed composite systems in Strongly damped reactions of
$^{238}$U+$^{238}$U, $^{232}$Th+$^{250}$Cf at $E_{cm}$=680-1880
MeV were studied based on the ImQMD model. One found that the
weakly repulsive entrance channel potential and strong dissipation
delay the re-separation time of a composite system, and a 15-20
MeV high Coulomb barrier at the surface of the single-particle
potential well of the composite system makes the excited unbound
protons still embedded in the potential well and to move in a
common mono-single-particle potential for a period of time. These
two effects restrains the quick decay of the composite system.
That study results in our interest for the incident-energy
dependence of lifetime of the composite system. We found that the
longest average lifetime for the composite system of
$^{238}$U+$^{238}$U could reach to over $\sim $1000 fm/$c$ at the
incident energy region 1000 to 1300 MeV. Recent study on the
incident-energy dependence of the lifetime of the transiently
formed giant composite system $^{238}$U+$^{238}$U by means of TDHF
calculations based on Skyrme energy density functional\cite{Ced09}
confirmed this results. Since the correlation (fluctuation) effect
is considered in the ImQMD model, it is able to calculate the mass
(charge) distribution of primary fragments in the
$^{238}$U+$^{238}$U reaction, in addition to study the properties
of the composite systems. The experiment for the reaction
$^{238}$U+$^{238}$U at energies close to the Coulomb barrier was
performed at GANIL and the mass distributions of products for the
reaction at several energies are available
now\cite{Heinz136,Golabek17}, which stimulates us to make further
study of the decay of the composite system of $^{238}$U+$^{238}$U.

In this work. we study the mass distribution of products in
$^{238}$U+$^{238}$U at 7 AMeV and then compare it with the
experimental data. Considering the extremely complexity of the
reaction process and saving computation time, we describe the
reaction process by a two step model, i.e. a dynamical reaction
process described by the ImQMD model followed by a statistical
decay process which is described by a statistical decay model.

The paper is organized as follows: In section II we will briefly
introduce the theoretical models. In section III and IV we present
the results of primary fragments and final products, respectively.
Finally, we will  give brief summary in section V.

\begin{center}
\bf{II. Theoretical Model}
\end{center}
Within this approach, the first step describes the formation and
re-separation process of the transiently formed composite systems
of $^{238}$U+$^{238}$U by means of the ImQMD model. The primary
fragments and fast particle emission are obtained at the end of
ImQMD calculations. The second step devotes to describe the decay
of the primary fragments by means of HIVAP incorporating with a
three Gaussian model for describing the mass distribution of
fission fragments. And finally the mass distribution of the
products is obtained.

\subsection{The ImQMD model}

Detailed description of the ImQMD model and its applications in
low energy heavy ion collisions can be found in Refs.
\cite{Wang05,Tian08,Wan02,Wan04}. Here, we only mention that in
this model the nuclear potential energy is an integration of the
potential energy density functional which reads
\begin{eqnarray}\label{Eqnarray5}
    V_{\texttt{loc}}~~=&&\frac{\alpha}{2}\frac{\rho ^{2}}{\rho _{0}}+\frac{\beta }{\gamma +1}%
    \frac{\rho ^{\gamma +1}}{\rho _{0}^{\gamma}}+\frac{g_{0}}{2\rho _{0}}\left(\nabla \rho \right)^{2} \nonumber\\
    &&+\frac{c_{s}}{2\rho_{0}}(\rho^{2}-\kappa_{s}(\nabla\rho)^{2})\delta^{2}+g_{\tau}\frac{\rho^{\eta+1}}{\rho_{0}^{\eta}},
    \end{eqnarray}
    where $\rho$, $\rho_{n}$, $\rho_{p}$ are the nucleon, neutron, and
    proton density, $\delta = (\rho_{n}-\rho_{p})/(\rho_{n}+\rho_{p})$
    is the isospin asymmetry.
The parameters in above expressions are given in the table
\ref{parameter}\cite{Wang05}.

\begin{table}[htbp]
\vspace{10mm}
\begin{center}
\begin{tabular}{ccccccccc}
 \hline
  $\alpha$(MeV) & $\beta$(MeV) & $\gamma$ & $g_0$(MeV$fm^{2}$) & $g_{\tau}$(MeV) & $\eta$ & $c_{s}$(MeV) & $\kappa_{s}$($fm^{2}$) & $\rho_0$($fm^{-3}$)\\
  \hline
  -356 & 303 & 7/6 & 7.0 & 12.5 & 2/3 & 32 & 0.08 & 0.165\\
 \hline
\end{tabular}
\end{center}
\caption{\label{parameter}the model parameters}
\end{table}
The Coulomb energy is also included in the Hamiltonian written as
a sum of the direct and the exchange contribution:
\begin{equation}
U_{Coul}=\frac{1}{2}\int \int \rho _{p}(\mathbf{r})\frac{e^{2}}{|\mathbf{r-r}%
^{\prime }|}\rho _{p}(\mathbf{r}^{\prime
})d\mathbf{r}d\mathbf{r}^{\prime }-e^{2}\frac{3}{4}\left(
\frac{3}{\pi }\right) ^{1/3}\int \rho _{p}^{4/3}d\mathbf{R}.
\label{17}
\end{equation}%
In the collision term, isospin-dependent nucleon-nucleon
scattering cross sections\cite{Cugnon96} are used and the Pauli
blocking effect is treated more strictly\cite{Zh06,LQF01}.

It is of crucial importance to make the initial nuclei in the real
ground state because considerable excitation of initial nuclei
will produce unreal particle emission and the residue with too
high excitation which will completely masks the real decay process
of residue. We check carefully not only the binding energy and the
root-mean-square radius of the initial nuclei but also their time
evolution. The average binding energy per nucleon of initial
nuclei is required to be $E_{g.s.}\pm0.1$ MeV, where $E_{g.s.}$ is
the binding energy of nuclei in ground state. It is required that
those initial nuclei with no spurious particle emission and their
properties such as binding energy and root-mean-square radius
being stable within 6000fm/c are taken to be as good initial
nuclei, and then are applied in the simulation of reaction
process. The deformation of the initial $^{238}$U
($\varepsilon$=0.24) is considered in the initial condition. In
the simulation of reactions, the initial orientations of two
deformed $^{238}$U are randomly taken. Fig.1 shows the time
evolution of the binding energy and root-mean-square radius of the
initial $^{238}$U.
\begin{figure}
\includegraphics[angle=270,scale=0.3]{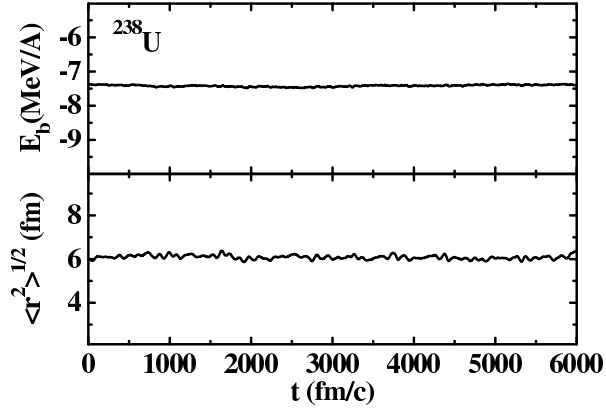}
\caption{The time evolution of the binding energy and
root-mean-square radius of the initial $^{238}$U.} \label{Fig.1}
\end{figure}

At the end of the ImQMD calculations, fragments are constructed by
means of the coalescence model widely used in the QMD
calculations. In this work only the primary fragments with mass
number larger than 50 are considered. Then, we calculate the total
energy of each excited fragment in its rest frame and its
excitation energy is obtained by subtracting the corresponding
ground state energy from the total energy of the excited fragment.

\subsection{The statistical decay model}

 The second step describes the
decay process of primary fragments by emission of neutron, proton
and $\alpha$ particle and fission. The statistical decay model
(HIVAP code)\cite{Reis85} incorporating a 3-Gaussian model for
mass distribution of fission fragments for fissile nuclei is used
to describe the decay process of primary fragments and mass
distribution of final products. In HIVAP, the survival probability
of an excited primary fragment is given by subsequent
de-excitation process leading to a given final evaporation-residue
nucleus in its ground state. Successive stages of a subsequent
de-excitation processes for primary fragment with mass A, charge Z
and excitation energy E are determined by branching ratios
expressed by relative partial decay widths for all possible decay
modes, $\Gamma_{i}(A,Z,E)/\Gamma_{tot}(A,Z,E)$, where
i=n,p,d,$\alpha$, etc., and $\Gamma_{tot}(A,Z,E)$ is the sum of
all particle decay widths $\Gamma_{i}(A,Z,E)$ and the fission
width $\Gamma_{f}(A,Z,E)$. All partial widths for emission of
light particles and fission for excited nuclei are calculated by
the HIVAP code.

The excited actinide and transactinide nuclei in primary fragments
and those produced in the de-excitation process undergo a fission.
The production probability of a fission fragment with mass number
$A_{1}$ is calculated as follows:
\begin{equation}
W_f(A_1)=\sum\limits_{A,Z,E}\frac{\Gamma_f(A,Z,E)}{\Gamma_{tot}(A,Z,E)}P(A_1,A,Z,E).
\end{equation}
Where the $P(A_{1},A,Z,E)$ is the production probability of a
fragment with mass number $A_{1}$ from a fission of the excited
nucleus with mass A, charge Z and excitation energy E. The
$P(A_{1},A,Z,E)$ is calculated based on an empirical three
gaussian model. It reads
\begin{equation}
P(A_{1},A,Z,E)=\sum\limits_{j=1}^{3}g^{(j)}(A_{1},A,Z,E)
\end{equation}
and
\begin{equation}
g^{(j)}(A_{1},A,Z,E)=\frac{P^{(j)}(A,Z,E)}{\sqrt{2\pi}\sigma^{(j)}(A,Z,E)}
\exp[-\frac{(A_{1}-A^{(j)}(A,Z,E))^{2}}{2(\sigma^{(j)}(A,Z,E))^{2}}],
\end{equation}

\begin{center}
                               j=1,2,3.
\end{center}

Where, the Gaussian distribution $g^{(j)}(A_{1},A,Z,E)$ represents
one of the components of the mass distribution of fission. Among
them, the $g^{(1)}(A_{1},A,Z,E)$ and $g^{(2)}(A_{1},A,Z,E)$
describe the asymmetric component of the mass distribution, and
$g^{(3)}(A_{1},A,Z,E)$ is for the symmetric component.
$P^{(j)}(A,Z,E)$, $\sigma^{(j)}(A,Z,E)$ and $A^{(j)}(A,Z,E)$ are
the parameters for 3 Gaussian distributions, which are the
function of mass number A, charge Z and excitation energy E of
fissile nucleus. The $P^{(j)}(A,Z,E)$ and $A^{(j)}(A,Z,E)$ obey
the following relations
\begin{equation}
P^{(1)}(A,Z,E)=(1-P^{(3)}(A,Z,E))\eta
\end{equation}
\begin{equation}
P^{(2)}(A,Z,E)=(1-P^{(3)}(A,Z,E))(1-\eta)
\end{equation}
\begin{equation}
A^{(1)}(A,Z,E)+A^{(2)}(A,Z,E)=A
\end{equation}
\begin{equation}
A^{(3)}(A,Z,E)=\frac{A}{2}
\end{equation}

Thus, only six parameters of $ P^{(3)}(A,Z,E)$, $\eta$,
$A^{(1)}(A,Z,E)$ and $\sigma^{(i)}$(i=1,2,3) are independent,
which need to be fixed according to available experimental data of
fission mass distributions in actinide and transactinide nuclei.

For fitting the parameters in the three Gaussian empirical formula
we collect available experimental data of fission mass
distributions\cite{Neuzil129,Benlliure628,Schmidt665,Gorodisskiy548,IAEA2008}
as many as possible. For the case of lack of experimental data the
interpolation or extrapolation method is employed. For $^{238}$U,
data for mass distributions of fission fragments at different
energies are available so we can obtain the energy dependence of
mass distribution of fission fragments through interpolation. But
for other fissile nuclei those data are relatively lack. For these
nuclei we suppose that they have similar energy dependence
behavior with those of $^{238}$U because the corresponding
theoretical study is also lack for these nuclei. This, of course,
will introduce a considerable approximation. However, in the
reaction considered in this work, the fission for excited
$^{238}$U is the most important one among all fissile nuclei and
we expect the approximation introduced in the energy dependence of
the mass distribution of fission fragments will not destroy the
final results. In Fig.2 and Fig.3 we show some examples of
calculated mass distributions of fission for different nuclei and
for different excitation energies and make comparison with
experimental data. The curves and dots are for calculated results
and data, respectively. From the figures we can see that the
empirical formula seems to be able to reproduce the available
experimental data and able to be used to calculate the mass
distributions of actinide and transactinide fragments.

\begin{figure}
\includegraphics[angle=270,scale=0.4]{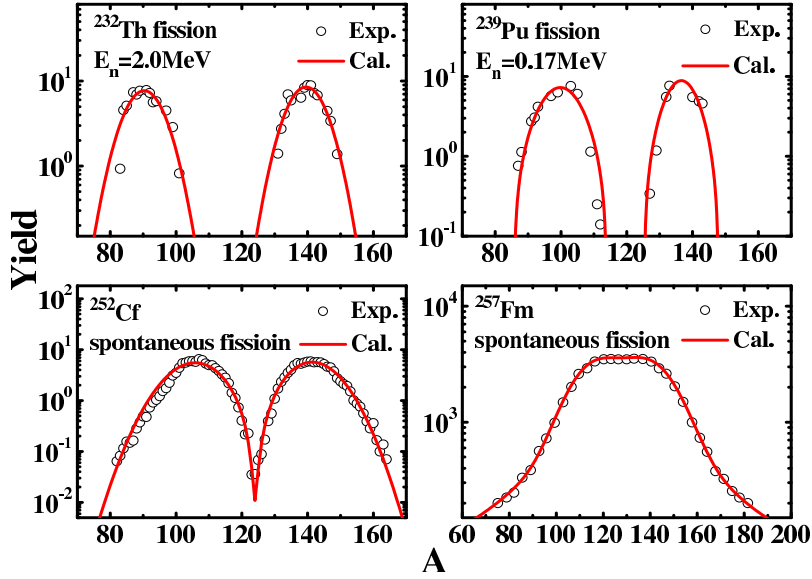}
\caption{(Color online) The mass distributions of fission for
$^{232}$Th, $^{239}$Pu, $^{252}$Cf and $^{257}$Fm. The
experimental data are taken
from\cite{Glen80,IAEA2008,Sch65,John71}} \label{Fig.2}
\end{figure}
\begin{figure}
\includegraphics[angle=270,scale=0.4]{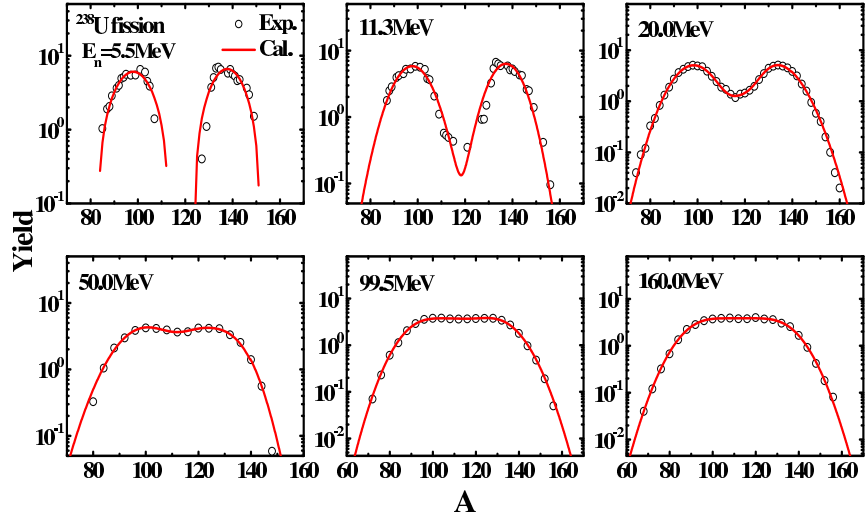}
\caption{(Color online) The mass distributions of fission for
$^{238}$U at different excitation energies. The experimental data
are taken from\cite{IAEA2008}} \label{Fig.3}
\end{figure}

In order to choose matching time t$_{S}$ of two models properly,
we investigate the decay process of the transiently formed
composite systems of $^{238}$U+$^{238}$U at the energy of 7.0
AMeV. FiG.4 shows the time dependence of the surviving probability
of fragments with Z$\geq$110. One can see from the figure that at
about t=500fm/c, two nuclei reach a touching configuration. After
about 1000fm/c the composite system begins to re-separate with a
very large decay rate and at about 3000 fm/c almost all composite
systems are separated. This process is described by the ImQMD
model. The separated fragments continue to decay with a much
smaller decay rate. This process is expected to be described by
the statistical decay model. Thus, we select the matching time of
two models to be 3000 fm/c. We have also tried other choices such
as t$_{S}$= 4000fm/c, 5000fm/c and 6000fm/c and we find there is
no change of final results. In the ImQMD calculations, 500 events
per impact parameter are performed.
\begin{figure}
\includegraphics[angle=270,scale=0.4]{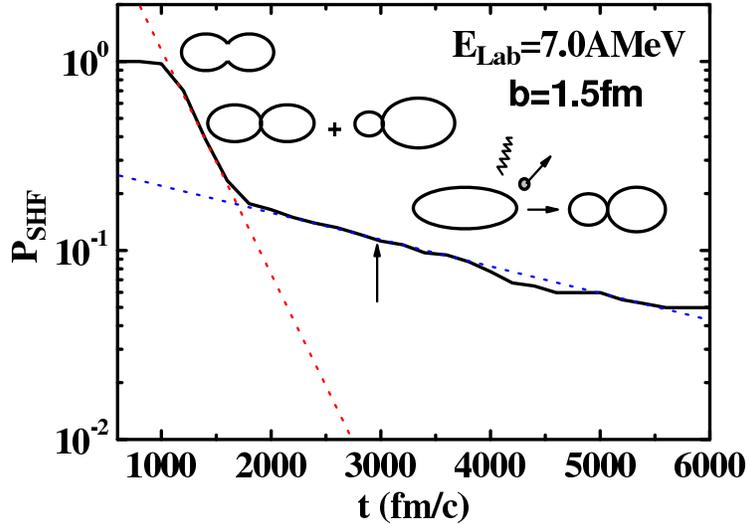}
\caption{(Color online) The time dependence of the surviving
probability of super-heavy fragments of Z$>$110.} \label{Fig.4}
\end{figure}

\begin{center}
\bf{III. The Distribution of Mass, Charge and Excitation Energy in
Primary Fragments}
\end{center}

In order to study the final mass distribution of the reaction
$^{238}$U+$^{238}$U, we first study the distribution of primary
fragments which are given at the end of ImQMD calculations. The
charge, mass and excitation energy distributions as well as the
angular distribution of primary fragments are obtained by the
ImQMD model calculations at time t=3000fm/c. The double
differential cross section of a primary fragment with charge Z,
mass A, excitation energy E and scattering angle $\theta$ is given
by:
\begin{eqnarray}
\frac{d^{2} \sigma_{pri}(Z,A,E,\theta)}{d\theta
 dE}=\int_{0}^{b_{max}}2\pi b
f(Z,A,E,\theta,b)db=\sum\limits_{i=1}^{i_{max}}2 \pi b_{i}\Delta b
f_{i}(Z,A,E,\theta,b_{i}), \label{13}
\end{eqnarray}
where $f_{i}(A,Z,E,\theta,b_{i})$ is the probability of producing
the primary fragments with charge Z, mass A, excitation energy E
and scattering angle $\theta$ under impact parameter $b_{i}$. The
maximum impact parameter $b_{max}$ is taken to be 14fm since there
are no inelastic scattering when b$>$14fm. The double differential
cross section for primary fragments will be used as input in the
second step for the calculations of final products in order to
compare with the measurement. Let us first study the charge and
mass distribution of primary fragments which is the integration of
double differential cross sections. Fig.5 (a) and (b) show the
charge and mass distribution of primary fragments(A $\geq$ 50) for
$^{238}$U+$^{238}$U at 7.0 AMeV, respectively. The sharp peak are
located at the uranium for both subfigures. The primary fragments
at the left side of the sharp peak stem from the re-separation of
the composite system and fast fission products of actinite and
transactinide fragments. The products at the right side of the
sharp peak correspond to transuranium nuclei.
\begin{figure}
\includegraphics[angle=270,scale=0.4]{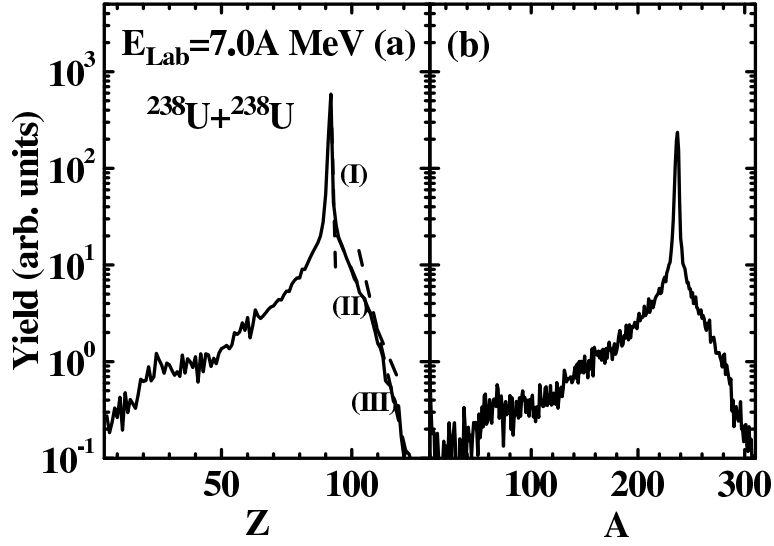}
\caption{(a)The charge and (b)the mass distribution of the primary
fragments of $^{238}$U+$^{238}$U at 7.0A MeV.} \label{Fig.5}
\end{figure}
The mass distributions of primary fragments at different impact
parameter $b$ are calculated in order to clarify the origin of the
fragments with different mass region. The results are shown in
Fig.6. Subfigure (a), (b), (c) and (d) are for the impact
parameter 0-4, 5-7, 8-10 and 11-14 fm, respectively. In central
collisions (see subfig.6(a)) the mass number distribution of
primary fragments extends to A=320 with a big asymmetric bump
around A=200-260, which means a large mass transfer between two
uranium nuclei happening in central collisions. At semi-central
collision (subfig.6(b)), the mass distribution becomes less wide
with a much shorter tail at right hand side. There are two peaks
appeared in the mass distribution with bigger one corresponding to
uranium and the smaller one originating from ternary
fission(occasionally from quaternary-fission)events in reaction
$^{238}$U+$^{238}$U. Clearly, in this case, very deep inelastic
reaction becomes the most important reaction mechanism. For the
peripheral collisions (subfig.6(c) and (d)) the mass distribution
of primary fragments shows a symmetric peak with a very less
variance. The reaction mechanism for peripheral reactions are of
the inelastic or elastic scattering between two uranium nuclei. In
order to understand the reaction mechanism and the mass
distribution of fragments evolving with impact parameters shown in
Fig.6 we present the average life-time of transiently formed
composite system for $^{238}$U+$^{238}$U at 7AMeV as function of
impact parameter in Fig.7.
 From this figure, one can see that the life-
time of composite system increases as impact parameter decreases.
In the central collisions, two uranium nuclei have longer
interaction time, stronger dissipation of collective motion and
thus have stronger mass transfer between them compared with larger
impact parameter cases. Therefore, the transuranium primary
fragments mainly come from the central and semi-central
collisions.
\begin{figure}
\includegraphics[angle=270,scale=0.4]{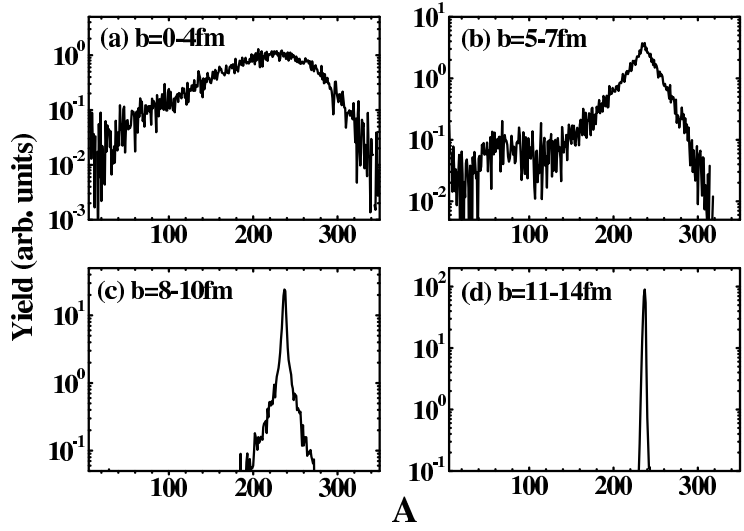}
\caption{The mass distributions of the primary fragments at
different impact parameter regions.} \label{Fig.6}
\end{figure}
\begin{figure}
\includegraphics[angle=270,scale=0.4]{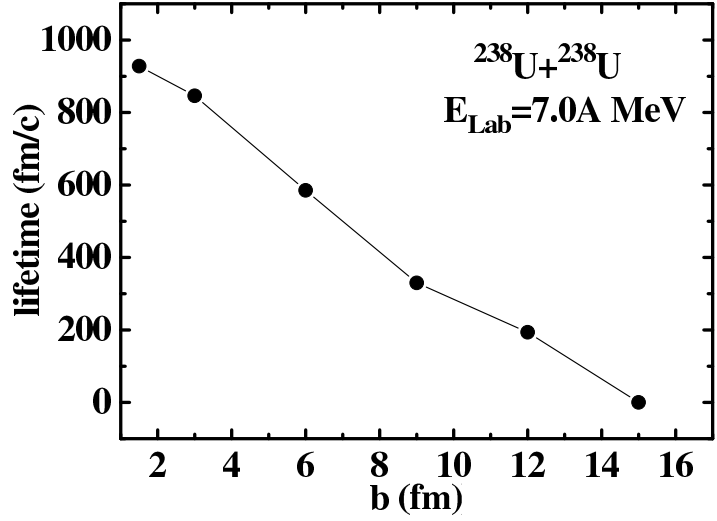}
\caption{The impact parameter dependence of the average life-time
for the composite system of $^{238}$U+$^{238}$U at 7 AMeV }
\label{Fig.7}
\end{figure}
Now we study the distribution of excitation energies of excited
fragments. Fig.8 and Fig.9 show the excitation energy
distributions for fragments with Z $\geq$ 100 and 90 $\leq$ Z
$\leq$ 94, respectively. As is mentioned above that the fragments
with Z $\geq$ 100 come from the large mass transfer reactions
which only happen in the central and semi-central collisions, the
results shown in Fig.8 are only for impact parameter b=0-4 and 5-7
fm. Fig.9 shows the results from deep inelastic scattering of
$^{238}$U+$^{238}$U. One sees from both Fig.8(a) and Fig.9(a) that
the primary fragments produced in central collisions are mostly
highly excited and for those fragments the survival probability
should be very low but still there is a tail extending to low
exciting energy, which may have certain but very small survival
probability. Whereas for the semi-central collision (see Fig.8(b)
and Fig.9(b)) the high excitation energy primary fragments
decreases and the portion of low energy primary fragments
increases, thus, it is expected that some of fragments with Z
$\geq$ 92 can be survival. In the peripheral collisions ( see
Fig.9(c) and (d))the excitation energies of primary fragments are
much lower compared with the central and semi-central cases and
the reason for it is of understandable.

\begin{figure}
\includegraphics[angle=270,scale=0.4]{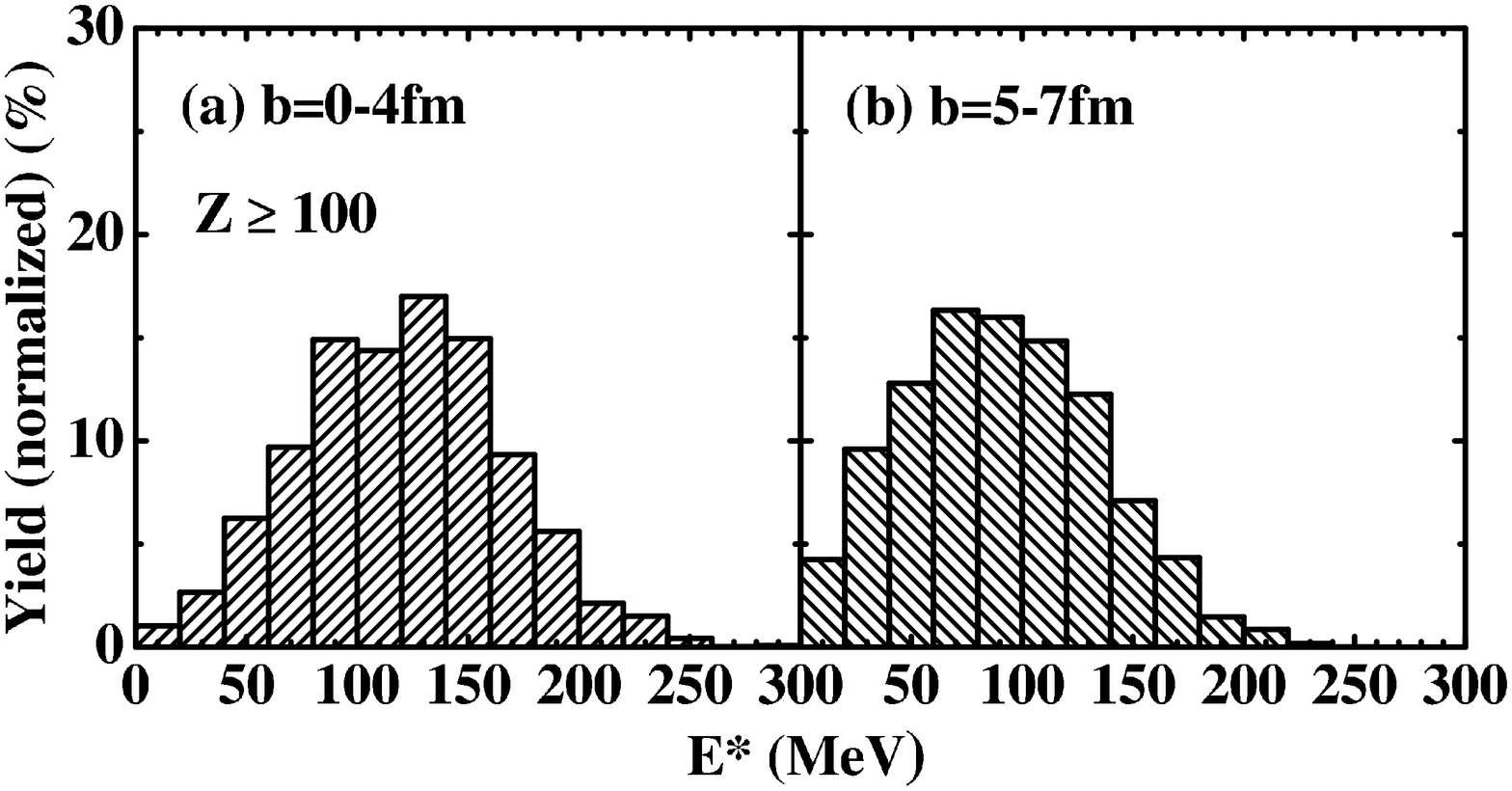}
\caption{The excitation energy distribution of the transuranic
fragments with $Z\geq100$ under condition of impact parameters (a)
b$\leq$4fm and (b) b=5-7fm, respectively.} \label{Fig.8}
\end{figure}

\begin{figure}
\includegraphics[angle=270,scale=0.4]{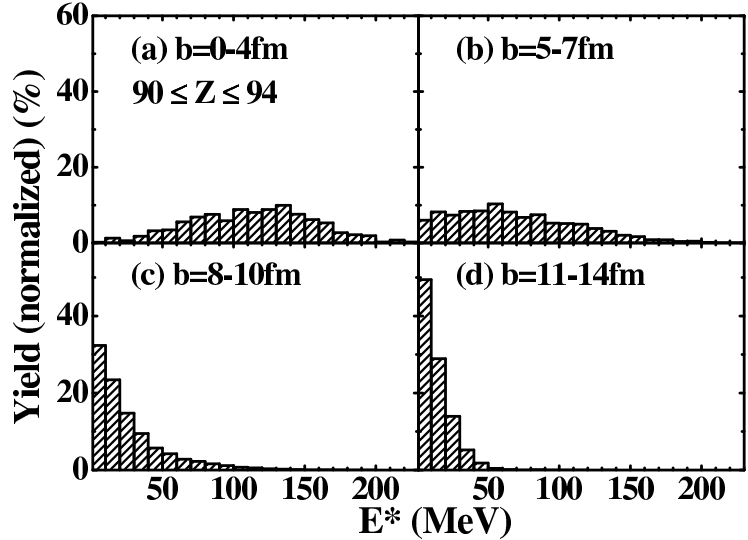}
\caption{The excitation energy distribution of the primary
fragments around uranium (90$\leq$Z$\leq$94) for impact parameters
(a) b$\leq$4fm, (b) b=5-7fm, (c) b=8-10fm and (d) b=11-13fm,
respectively.} \label{Fig.9}
\end{figure}

\newpage

\begin{center}
\bf{IV. Mass Distribution of Final Products}
\end{center}
From the ImQMD model calculation we obtain the distributions of
charges, masses and excitation energies for all produced primary
fragments in $^{238}$U+$^{238}$U collisions. These primary
fragments will de-excite through emitting light particles,
$\gamma$-rays and fission. The decay process and the final
products are described by the statistical evaporation model (HIVAP
code) incorporated the three Gaussian fission model described in
Section II. Based on the model, the mass distribution of final
products for $^{238}$U+$^{238}$U at the incident energy of 7.0
AMeV can be calculated. In Fig.10, we show the calculated results
of final products at 4 impact parameter regions of 0-4, 5-7, 8-10
and 11-14 fm. For central collisions( see Fig.10(a)), the
re-separation primary fragments of $^{238}$U+$^{238}$U systems
carry high excitation energies, the most part of them undergoes
symmetric fission and thus a single bump of mass yield is found at
around mass number 120. The rest of fragments not undergoing
fission will evaporate particles and their residues finally form a
shoulder in the mass distribution around Pb, which is due to
strong shell effect for those nuclei around Pb. The yields for
transuranic fragments decrease rapidly as mass increase, which is
due to the high excitation energy of primary fragments in central
collisions as seen from Fig.8(a). Here we should mention that the
yields of the transuranic nuclei is not so certain because the
fission barrier and the fission width for super-heavy nuclei and
the transuranic nuclei are largely uncertain. For semi-central
collisions, i.e. in the impact parameter region of 5-7 fm (see
Fig.10(b)), the excitation energies carried by primary fragments
are much less than those in the central collisions, so there
appears a broad bump at mass number region of 80$\leq$A$\leq$170
which is the superposition of symmetric and asymmetric fission.
There appears another small bump centered at Uranium(A $\approx$
230). The shallow valley between two bumps means that the yields
of nuclei around Pb is still considerable. Here we notice that the
yields of transuranic nuclei is relatively higher than those in
central collisions, which is because the excitation energies of
primary fragments are much lower than those in central collisions.
For peripheral collisions (see Fig.10(c) and (d)), elastic or
inelastic scattering play a dominant role and the behavior of low
energy fission of actinide nuclei is shown. The small shoulder
around Pb seems to appear for impact parameters b=8-10fm(see
Fig.10(c)).

In order to make comparison with experimental measurement we have
to make selection of scattering angle to fit the angle cut in
experimental data, i.e. only fragments with the scattering angles
of 56$^{o}$ $\leq \theta \leq$ 84$^{o}$ and 96$^{o}$ $\leq \theta
\leq$ 124$^{o}$ in the center of mass frame are
selected\cite{Golabek17}. In the calculations, we assume that the
scattering angle of residue of the primary fragment which
undergoes emission of light charged particles is the same as the
fragment itself. This assumption is roughly reasonable since the
mass of residue is much larger than that of emitted light
particles. For fragments from fission, we assume that the outgoing
angle of one fragment is randomly distributed in the rest frame of
the fissioning nucleus and the outgoing angle of the other one is
then obtained by momentum conservation. Finally, we obtain the
mass distribution of the final products with the same scattering
angle cut as that in the experiment. The results are shown by open
triangles in Fig.11. The experimental mass spectra from
\cite{Golabek17} are also indicated by solid squares, open
squares, solid circles, open circles and solid triangles for
incident energies of 6.09, 6.49, 6.91, 7.10 and 7.35 AMeV,
respectively, in Fig.11. From the figure we find that the behavior
of the calculated mass distribution at 7.0 AMeV is generally in
agreement with the data at incident energy 7.10 AMeV except the
yields at mass region from 170 to 210 to be overestimated compared
with experimental data. The following most important features of
mass distribution can be deduced: (1) A dominant peak around
uranium is observed, which can be attributed to the contribution
of the reactions with large impact parameters seeing from Fig.10;
(2) The steep decreasing yield above U with the increase of the
mass number is appeared. The products at this mass region stem
from large mass transfer in small impact parameter reactions; (3)A
small shoulder can be seen in the distribution of the products
around Pb, compared with the products with mass near and smaller
than Uranium for which the yields decrease exponentially as mass
decreases. The appearing of small shoulder around Pb is due to the
very high fission barrier around Pb. The central, semi-central
collisions and even reactions with b=8-10fm contribute to the
shoulder around Pb region; (4) In the region below A$\approx$190,
the double bump distribution is observed. These products are from
the fission of actinide and transuranium nuclei obviously, which
are superposition of symmetric and asymmetric fission.

\begin{figure}
\includegraphics[angle=270,scale=0.4]{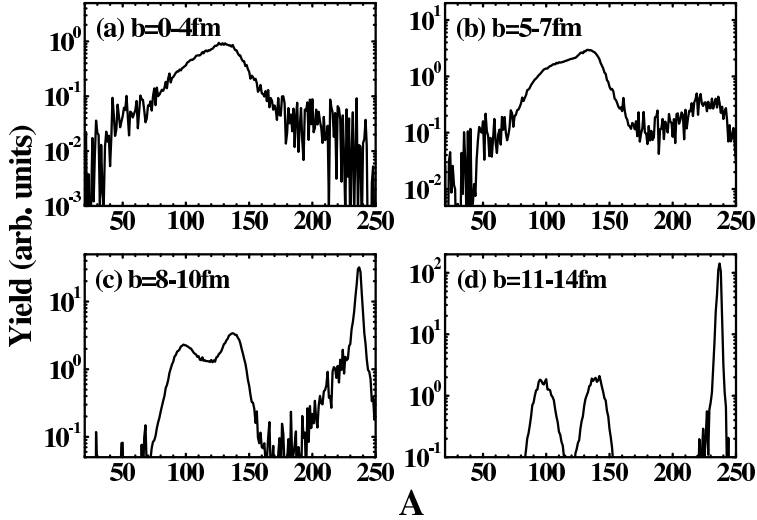}
\caption{The mass distribution of the products in $^{238}$U+
$^{238}$U at different impact parameter regions.} \label{Fig.10}
\end{figure}

\begin{figure}
\includegraphics[angle=270,scale=0.4]{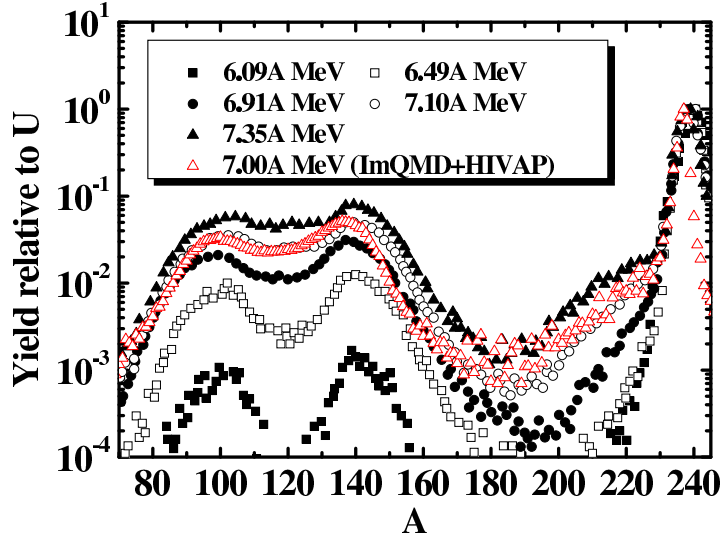}
\caption{(Color online) The mass distribution of the products of
reaction $^{238}$U+$^{238}$U. The experimental data are from
\cite{Golabek17}.} \label{Fig.11}
\end{figure}

\begin{center}
\bf{V. SUMMARY}
\end{center}

In this paper, we apply the microscopic transport model, namely
the ImQMD model incorporating the statistical decay model and
empirical three Gaussian fission model to study the reaction
mechanism and the mass distribution of products in the reaction
$^{238}$U+$^{238}$U at the incident energy of 7.0 AMeV. The mass,
charge and excitation energy distributions of primary fragments
are calculated within the ImQMD model and  the de-excitation
process of those primary fragments is studied by using the
statistical-evaporation model (HIVAP code). The impact parameter
dependence of the mass distribution of primary fragments and final
products are analyzed, from which the origin of products at
different mass region can be understood. Finally, the mass
distribution of final products in $^{238}$U+$^{238}$U collisions
with scattering angle cut is calculated in the first time and
compared with recent experimental data. The main features of
experimental mass distribution are reproduced, those are: (1)A
dominant peak around the uranium nuclei is observed, which
corresponds to elastic and quasi-elastic reaction products; (2)The
yields of the transuranium nuclei decrease rapidly with increase
of the mass A ; (3)A small shoulder can be seen in the mass
distribution of the products around Pb on the background of
products for which their yields decrease with their mass deviating
from uranium exponentially.  Those products are the residues of
primary fragments surviving from multiple-particle evaporation;
(4)In the region below A$\approx$200, the double hump mass
distribution is observed, which are the fission products from
superposition of symmetric and asymmetric fission and mainly come
from the fission of nuclei around uranium and transuranium
fragments at high and low excitation energies. The main
discrepancy of our calculated results with experimental data is
overestimate of mass yields in region of 170-200 and underestimate
of mass yield of transuranium nuclei, which mainly come from the
calculation of fission mass distribution for actinide and
transuranium nuclei and fission width at high excitation energies.
Further study is still under a way.
\begin{center}
{\bf ACKNOWLEDGMENTS}
\end{center}

This work is supported by the National Natural Science Foundation
of China Nos. 10675172, 10875031, National Basic Research Program
of China No.2007CB209900.

\end{document}